\documentclass[eqsecnum,floats,aps,nofootinbib,showpacs]{revtex4}
\usepackage{latexsym}
\usepackage{amssymb}
\usepackage{ifpdf}
\ifpdf
\usepackage[pdftex]{hyperref}
\else
\usepackage{hyperref}
\fi

\newcommand{\LL}{\mathcal{L}}
\renewcommand{\d}{\mathrm{d}}
\renewcommand{\l}{\left(}
\renewcommand{\r}{\right)}
\usepackage{amsmath, amsthm, amssymb}

\def\be{\begin{equation}}
\def\ee{\end{equation}}
\def\beq{\begin{equation*}}
\def\eeq{\end{equation*}}
\def\ba{\begin{aligned}}
\def\ea{\end{aligned}}

\def\p{\partial}
\def\ov{\overline}
\def\w{\wedge}

\begin{document}
\title{Nonuniqueness of gravity--induced fermion interaction in the Einstein--Cartan theory}

\author {Marcin Ka\'zmierczak}
\email{marcin.kazmierczak@fuw.edu.pl}
\affiliation{Institute of Theoretical Physics, University of Warsaw
  ul. Ho\.{z}a 69, 00-681 Warszawa, Poland}

\begin{abstract}
The problem of nonuniqueness of minimal coupling procedure for
Einstein--Cartan (EC) gravity with matter is investigated. It is shown
that the predictions of the theory of gravity with fermionic matter can
radically change if the freedom of the addition of a divergence to the
flat space matter Lagrangean density is exploited. The well--known
gravity--induced four--fermion interaction is shown to reveal
unexpected features. The solution to the problem of nonuniqueness of
minimal coupling of EC gravity is argued to be necessary
in order for the theory to produce definite predictions. In
particular, the EC theory with fermions is shown to be
indistinguishable from usual General Relativity on the effective
level, if the flat space fermionic Lagrangean is appropriately
chosen. Hence, the solution to the problem of nonuniqueness of minimal coupling
procedure is argued to
be necessary if EC theory is to be experimentally
verifiable. It could also enable experimental tests of theories based on EC, such as loop
approach to quantisation of gravitational field. Some ideas of
how the arbitrariness incorporated in EC theory could be restricted or
even eliminated are presented.
\end{abstract}
 
\maketitle

\section{Introduction}
\label{section1}

Einstein--Cartan theory (EC) is acknowledged as a viable alternative
for General Relativity (GR), confirmed by all available experimental
data. For an exhaustive review of the theory, see \cite{HHKN}. For a
mathematically rigorous formulation in terms of tensor valued
differential forms, see \cite{T1}. If coupled to fermions, the EC theory is claimed to differ from
General Relativity by the presence of a gravity--induced four--fermion
interaction. The
effective action obtained by integrating out the connection contains
an  additional term, when compared to GR, which is proportional to the
square of an axial fermion current. The equation for Dirac
bispinor field is nonlinear, even in the limit of the space--time
metric being Minkowski's flat one. This nonlinearity can be
interpreted as describing interaction between fermions. Some
interesting properties of this interaction have been studied \cite{Ker}\cite{Rumpf}\cite{CI}. It is
generally believed that, although in principle it is measurable, its effects
cannot be measured in practice due to the
smallness of coupling constant appearing in front of a new term in the
effective action. 

The interest in gravity--induced four--fermion interaction increased in
the last few years because of the development of canonical approach
to the quantisation of gravity. Since the introduction by Ashtekar of a new
formalism for complex General Relativity (GR) \cite{A1}\cite{A2}, reducing
constraints of the theory to the
polynomial form, many steps forward in the program of quantisation 
have been made. A new formalized treatment of nonperturbative
canonical gravity has emerged, known as
{\it Loop Quantum Gravity} \cite{AL}. In order to avoid difficulties
concerning reality conditions, necessary in Ashtekar complex
approach, Barbero \cite{Bar} proposed a real alternative. The
relation between these two approaches was then clarified by
Immirzi \cite{Imm} and Holst \cite{Hol}. 
It appears that adding a
new term to the standard Palatini action of General Relativity allows a
unified treatment of them. If the multiplicative parameter $\beta$ is
introduced in front of the  new term, the theory reduces to that of Ashtekar and Barbero for
$\beta=\pm i$ and $\beta=\pm 1$ respectively. The new constant
$\beta$ is called the Immirzi parameter. In the case of absence of torsion
generating matter, the additional term, called the Holst term, does
not influence field equations, as it vanishes on account of Bianchi
identity. Hence, the Immirzi parameter drops out from the classical
theory but appears to play an important role in quantum theory as it
enters the spectra of area and volume operators \cite{RT}. This
allowed for the establishing of theoretical bounds on possible values of the Immirzi
parameter by black hole entropy calculations and comparison with the
Bekenstein--Hawking formula \cite{DL}. A precise value of
the parameter was given shortly after \cite{Meiss}. Then it was noted that
even in the classical theory of gravity, the new Holst term does
influence field equations when fermions are minimally coupled
\cite{PR}. This initiated discussion on the role played by
the Immirzi parameter in classical gravity with fermions
\cite{FMT}\cite{Ran}\cite{Merc}\cite{BD}\cite{Alex}. As originally
observed in \cite{PR}, the Immirzi parameter enters the coupling constant
in front of the four--fermion interaction term of EC gravity. It has
been concluded that measuring the strength of this interaction can
provide a tool to estimate the value of the Immirzi parameter
independently from the quantum theory of gravity. Although the subsequent
investigations \cite{Merc}\cite{Alex} showed that the Immirzi
parameter can be ``hidden'' in the parameters of more general,
non--minimal coupling procedures, it should be stressed that the
minimal coupling scheme has been historically successful in constructing models, which could
withstand the rigors of experimental testing whenever such tests were
feasible. The experimental successes of the
standard model of particle physics and general theory of
relativity seem to support the minimal approach. Indeed, the Yang--Mills theories, which constitute the
formal basis for the standard model, employ minimal coupling scheme on
the fundamental level. The necessity of using non--minimal couplings
when describing effectively composed objects does not hold much relevance
as long as we aim to incorporate elementary point--like fermions (quarks and
leptons) into the theory of gravity. As is well known, the EC
gravity can be formulated as a gauge theory of Yang-Mills type for the
Poincar{\'e} group. Hence, one could hope that the application of
minimal coupling would lead to the physically relevant model in this case as
well. According to this viewpoint, special importance should be
attached to the original analysis of \cite{PR},
rather than to the later analyses employing
non--minimal couplings.

However, there is an important issue which seems to have been overseen
in most considerations concerning predictions of EC gravity with
fermions. The standard minimal coupling procedure (MCP) that simply
means converting all partial derivatives in flat space matter
Lagrangean into covariant ones and applying metric volume element to
construct Lagrangean four--form is not unique in the case of torsion
connections. Equivalent flat Lagrangeans (generating the same flat
space field equations) give rise to curved theories which are not in
general equivalent. 
That issue has been discussed since the very beginning of
gauge formulation of gravity \cite{Kib1}. The nature of the
problem is recalled in  Section \ref{section2}. One approach to solve it, reconsidered
here in Section \ref{EMS}, is
to set up the procedure for choosing the `appropriate' flat space Lagrangean
from the whole class of equivalent ones. The possibility of using
Noether theorem to achieve that purpose is investigated.
 More radical
solution would be to modify MCP itself to make
it give equivalent results for equivalent Lagrangeans. Such
modification was proposed by Saa in \cite{Saa1}\cite{Saa2} leading to
interesting effects, such as propagating torsion or coupling gauge
fields to torsion without breaking gauge symmetry. Although Saa's idea
provides a very interesting solution to the problem, it results in
significant departures from standard GR, which are not certain to
withstand the confrontation with observable data \cite{BFY}\cite{FY} without
some assumptions of rather artificial nature, such as demanding
a priori that part of the torsion tensor vanish \cite{RMAS}. 

In this paper we wish to argue that solving the MCP nonuniqueness
problem is crucial if EC gravity is to produce any nontrivial definite
predictions. In Section \ref{ECgravity}, we briefly recall the
formalism of EC theory and comment on compatibility of different
possible definitions of energy--momentum and spin density tensors. In
Section \ref{EMS}, we impose some reasonable restrictions on flat space
fermionic Lagrangeans which leaves us with two--parameter family. We
also give plausible arguments in favour of one particular choice. Then
in Section \ref{effective} we rederive the effective action and modified Dirac
equation for this family. We show that the above mentioned freedom can lead to the change of coupling constant
of axial--axial four--fermion interaction, as well as appearance of a
new vector--vector and even parity breaking axial--vector interaction
\footnote{Although it was mentioned in \cite{HD} that an
  axial--vector interaction might appear if flat space fermion
  Lagrangean were appropriately chosen, no explanation of the nature
  of this alternative flat Lagrangean can be found, nor were the reasons
  for not treating it seriously given. No mention of
  possibility of vector--vector interaction, or the change of coupling
  constant appeared, either.}. We show that even disregarding the technical
limitations one cannot distinguish experimentally between
EC gravity with fermions and standard torsionless GR treatment before
the nonuniqueness problem of EC gravity with matter is solved, unless
torsion is directly measurable. Finally, we try to understand the
physical nature of the new interaction by employing the background
field approximation. Our results show that the earlier
statements \cite{Ker}\cite{Rumpf} of universality (independence of
matter type) of interaction remain valid, whereas the question of whether the interaction is
repulsive or attractive for aligned and antiparallel spins remains
open, the answer being dependent on the values of parameters of our generalized model.

\section{Nonuniqueness of minimal coupling procedure}\label{section2}
A classical field theory in flat Minkowski space is defined by the
action functional
\begin{equation*}
S=\int \mathfrak{L} \ ,
\end{equation*}
where $\mathcal{L}$ is a Lagrangean density and
$\mathfrak{L}=\mathcal{L}\, \d^4 x=\mathcal{L}\,\d x^0\wedge\d
x^1\wedge\d x^2\wedge\d x^3$ a Lagrangean four--form. 
It is well known that the addition of a divergence of a vector field $V$ to
$\mathcal{L}$ changes $\mathfrak{L}$ by a differential
\be\label{differential}
\partial_{\mu}V^{\mu}\, \d^4 x=L_V\,\d^4 x=\d (V \lrcorner \, \d^4 x) \ ,
\ee
where $L$ denotes Lie derivative and $\lrcorner$ the internal
product. Thus, such a transformation does not change field equations
generated by $S$. In order to proceed from Minkowski space to the general
Riemann--Cartan (RC) manifold with metric $g_{\mu\nu}$ and the metric
compatible connection
$\nabla$ (not necessarily torsion--free) \footnote{For more general
  considerations concerning not necessarily metric connections see \cite{RMAS}.}, we can apply MCP
\be\label{MCP}
\int \mathcal{L}(\phi,\partial_{\mu}\phi,\dots )\,\d^4 x \
\longrightarrow \ \int \mathcal{L}(\phi,\nabla_{\mu}\phi,\dots
)\,\epsilon \ ,
\ee
where $\epsilon =\sqrt{-g}\,\d^4 x$ is the metric volume form, $g$ being the
determinant of a matrix of components $g_{\mu\nu}$ of the metric
tensor in the basis $\partial_{\mu}$, and $\phi$ represents fields of
the theory. Dots in (\ref{MCP}) correspond
to the possibility of $\mathcal{L}$ to depend on higher derivatives of
fields. Had we used the modified flat space Lagrangean
$\mathcal{L}+\partial_{\mu}V^{\mu}$, we would have obtained different
Lagrangean four--form on RC manifold, the difference being
\be\label{TV}
\nabla_{\mu}V^{\mu}\epsilon={\stackrel{\circ}{\nabla}}_{\mu}V^{\mu}\epsilon-T_{\mu}V^{\mu}\epsilon
\ ,
\ee
where $\stackrel{\circ}{\nabla}$ is the torsion--free Levi--Civita connection and
$T_{\mu}={T^{\nu}}_{\mu\nu}$ the torsion trace vector. The first term
in (\ref{TV}) is a differential,
${\stackrel{\circ}{\nabla }}_{\mu}V^{\mu}\epsilon=\d (V \lrcorner \, \epsilon)$,
whereas the second is not. Hence, the equivalent flat Lagrangeans yield
nonequivalent theories on RC space. 

One could hope differential forms formalism would fix the problem and
argue that the last expression of
(\ref{differential}), rather than the first, should be adopted to curved space. Then, $\d (V \lrcorner \,
\d^4 x)$ would transform into $\d (V \lrcorner \, \epsilon)$, which is
again a differential. However, this is not a good solution, since decomposition of a given Lagrangean four--form 
$\mathcal{L}_1\,\d^4 x$ to the sum of another Lagrangean four--form $\mathcal{L}_2\,\d^4 x$ and
the term $\d (V\lrcorner \,\d^4 x)$ is by no means unique. We should
rather use the identity $\d (V \lrcorner \, \d^4 x)=-\star \d x_{\mu}
\wedge \d V^{\mu}$, where $\star$ is a hodge star (see Section \ref{A1}), and minimally couple gravity by the passage
$\d V^{\mu}\longrightarrow DV^{\mu}=\d V^{\mu}+{\omega^{\mu}}_{\nu}
V^{\nu}$ (where ${\omega^{\mu}}_{\nu}$ are connection one--forms) and by the
change of a hodge star of flat Minkowski metric to the one of curved
metric on the finall manifold, but this would give the result identical to (\ref{TV}).

\section{Energy--momentum and spin tensors for equivalent Lagrangeans}\label{EMS}
In this section we will investigate transformation properties of
Noether currents of physical interest under the addition of a divergence
to the Lagrangean. The leading idea is that demanding these currents
to have the most `reasonable' form may help us restrict the class
of equivalent Minkowski space Lagrangeans and thus limit the
number of nonequivalent theories on RC space--time which are worth
further considerations. 
Let $\mathcal{L}(\phi^A,\partial_{\mu}\phi^A)$ be a Lagrangean density
of a field theory in Minkowski space. The invariance of
$\mathcal{L}$ under the global action of a Lie group of transformations $x^{\mu}\rightarrow x^{\mu}+\delta x^{\mu}$ ,
$\phi^A\rightarrow \phi^A+\delta \phi^A$ which do not change the
volume--form $\d^4 x$ of Minkowski metric implies that the Noether current
\[
j^{\mu}=-{t_{\nu}}^{\mu}\delta x^{\nu}+\frac{\partial\mathcal{L}} 
{\partial( \partial_{\mu} \phi^A )} \delta\phi^A
\]
is conserved, i.e. $\partial_{\mu}j^{\mu}=0$, if field equations are
satisfied. Here
\be\label{t}
t_{\mu\nu}:=\frac{\partial\mathcal{L}}{\partial (\partial^{\nu}\phi^A)}\partial_{\mu}\phi^A-\eta_{\mu\nu}\mathcal{L} \ .
\ee
is the {\it canonical energy--momentum tensor}, constituting the set
of currents which are conserved due to the symmetry of $\mathcal{L}$ under space--time
translations $x^{\mu}\rightarrow x^{\mu}+a^{\mu}$. 
In the case of $\mathcal{L}$ being invariant under proper Lorentz transformations (all
Minkowski space Lagrangean densities considered in this paper posses
that property), the corresponding conserved currents comprise an {\it
  angular momentum tensor} $\mathcal{M}^{\nu\alpha\mu}$
\[
\mathcal{M}^{\nu\alpha\mu}\varepsilon_{\nu\alpha}=2j^{\mu}=
(x^{\nu}t^{\alpha\mu}-x^{\alpha}t^{\nu\mu}+ 
S^{\nu\alpha\mu}) \varepsilon_{\nu\alpha}
\]
where $\varepsilon_{\mu\nu}$ are parameters of a Lorentz transformation
(${\Lambda(\varepsilon)^{\mu}}_{\nu}\equiv{\delta^{\mu}}_{\nu}+{\varepsilon^{\mu}}_{\nu}$ for small
$\varepsilon$). Here the {\it spin density tensor} $S^{\nu\alpha\mu}$ is a part of
$\mathcal{M}^{\nu\alpha\mu}$ which depends on transformation properties of
fields $\phi^A$. If $\delta\phi^A$ is
known, it can be computed from 
\be\label{S}
S^{\nu\alpha\mu}\varepsilon_{\nu\alpha}= 
2 \frac{\partial\mathcal{L}} 
{\partial(\partial_{\mu}\phi^A)}\delta\phi^A \ .
\ee
Apart from obvious scaling freedom, we can construct new conserved
currents from a given one via the transformation
\be\label{f}
j'^{\mu}=j^{\mu}+\partial_{\nu}f^{\mu\nu} \ ,
\ee 
where $f^{\mu\nu}=-f^{\nu\mu}$. In addition to giving another
conserved current, transformation (\ref{f}) does not change integrated
charges $Q=\int_{t=const.}j^0 \d^3\,\vec x$, if $f^{\mu\nu}$ vanishes sufficiently fast at spatial
infinity.

Let us now consider two Lagrangean densities differing by a divergence
of a vector field $V^{\mu}(\phi^A)$ (we wish $V^{\mu}$ not to depend
on derivatives of $\phi^A$ in order for both Lagrangeans to depend on
first derivatives only)
\be\label{Lch}
\mathcal{L}-\mathcal{L}'=\partial_{\mu}V^{\mu}=\frac{\partial
  V^{\mu}}{\partial \phi^A}\partial_{\mu}\phi^A \ .
\ee
Here, $V$ is required to transform as a vector under proper Lorentz
transformations: if $\phi^A\rightarrow \phi'^A$ represents the
action of a relevant representation of a proper Lorentz group in the
space of fields, we have $V^{\mu}\l{\phi^A}\r\rightarrow
V^{\mu}\l{\phi'^A}\r={\Lambda^{\mu}}_{\nu}V^{\nu}\l{\phi^A}\r$. Hence,
$\partial_{\mu}V^{\mu}$ is a Lorentz scalar and $\mathcal{L}'$
is a Lorentz scalar (if $\mathcal{L}$ is a scalar). All Lagrangean densities considered by us are also
required to be real, which implies reality of $V$.
The difference in energy--momentum and spin tensors corresponding to
(\ref{Lch}) will be
\be\label{tSch}
{{t'}_{\mu}}^{\nu}-{{t}_{\mu}}^{\nu}=\partial_{\rho}{f_{\mu}}^{\nu\rho} \ , \qquad 
({S'}^{\alpha\beta\mu}-S^{\alpha\beta\mu}) \, \varepsilon_{\alpha\beta}=
2\frac{\partial V^{\mu}}{\partial\phi^A}\delta\phi^A \ ,
\ee 
where
${f_{\mu}}^{\nu\rho}=\delta^{\rho}_{\mu}V^{\nu}-\delta^{\nu}_{\mu}V^{\rho}$.
Since ${f_{\mu}}^{\nu\rho}=-{f_{\mu}}^{\rho\nu}$, the change of
${t_{\mu}}^{\nu}$ corresponds to the usual freedom (\ref{f}) left by Noether
procedure and integrated energy and momenta do not transform. Hence,
the energy--momentum tensor cannot help us choose among Lagrangeans
differing by a divergence. Let us then focus our attention on spin
tensor. For the purposes of this paper, we will confine ourselves to the
case of a Dirac bispinor field $\psi$ and the vector field of the form
\be\label{VB}
V^{\mu}=\ov{\psi}B^{\mu}\psi \ ,
\ee
where $B^{\mu}=a\gamma^{\mu}+b\gamma^{\mu}\gamma^5$ for some real
numbers $a$ and $b$ (recall that $V$
is required to transform as a vector under proper Lorentz
transformations). Here $\gamma^{\mu}$ are the Dirac matrixes obeying 
$\gamma^{\mu}\gamma^{\nu}+\gamma^{\nu}\gamma^{\mu}=2\eta^{\mu\nu}$, 
$\gamma^5:=-i\gamma^0\gamma^1\gamma^2\gamma^3$ and
$\ov{\psi}:={\psi}^{\dagger}\gamma^0$, where ${\psi}^{\dagger}$ is
hermitian conjugation of a column matrix. Hence, 
\be\label{Vab}
V^{\mu}=aJ_{(V)}^{\mu}+bJ_{(A)}^{\mu} \ ,
\ee
where $J_{(V)}^{\mu}=\ov{\psi}\gamma^{\mu}\psi$,
$J_{(A)}^{\mu}=\ov{\psi}\gamma^{\mu}\gamma^5\psi$ denote Dirac vector and
axial current. For the relevant
representation of the Lorentz group
\be\label{repr}
S\left({\Lambda (\varepsilon)}\right)=\exp
\left({-\frac{i}{4}\varepsilon_{\alpha\beta} {\Sigma}^{\alpha\beta}}\right)
 \ , \qquad
 \Sigma^{\alpha\beta}=\frac{i}{2}[\gamma^{\alpha},\gamma^{\beta}] \ , \qquad
\psi\rightarrow S\left({\Lambda }\right)\psi \ ,
\ee
we have
\be\label{psich}
\delta\psi=-\frac{i}{4} 
\varepsilon_{\alpha\beta} {\Sigma}^{\alpha\beta}\psi \ , \qquad 
\delta\ov{\psi}=\frac{i}{4}\varepsilon_{\alpha\beta}\ov{\psi}\Sigma^{\alpha\beta} \ ,
\ee
which yields
\be\label{Sch}
{S'}^{\alpha\beta\mu}-S^{\alpha\beta\mu}=
\frac{i}{2}\ov{\psi} [\Sigma^{\alpha\beta},B^{\mu}]\psi \; ,
\ee
on account of (\ref{tSch}). One can observe that the expression
vanishes for densities of spatial components of spin:
${S' \, }^{ij0}=S^{ij0}$. Hence, for all Lagrangeans differing
by divergence from the standard one 
\be\label{LF0}
\mathcal{L}_{F0}=\frac{i}{2}\left({\ov{\psi}\gamma^{\mu}\partial_{\mu}\psi-\partial_{\mu}\ov{\psi}\gamma^{\mu}\psi}\right)-
m\ov{\psi}\psi
\ee
(throughout the paper we use the $c=\hbar =1$ units) we have
\[
S^{ij0}=\frac{1}{4}\ov{\psi}\left\{{\Sigma^{ij},\gamma^0}\right\}\psi=\frac{1}{2}\psi^{\dagger}\Sigma^{ij}\psi
\ .
\]
If integrated over the space, this yields an expected value at a state
represented by a wave function $\psi(x)$ of the correct spin operator (first quantisation interpretation
of $\psi$ is applied). Hence, densities of
space--space spin components cannot be used to choose appropriate
Lagrangean. As far as time--space components are concerned, the
situation is more interesting as their densities do transform according
to
\[
{S'}^{0j0}-S^{0j0}=\ov{\psi}B^j\psi \ .
\]
We can see that corresponding integrated charges are also
different. If we were able to measure them, we could choose between
$\mathcal{L}$ and $\mathcal{L}'$. Unfortunately, as pointed out by Kibble
\cite{Kib1}, it is not clear whether any physical significance should
be attached to the separation of time--space components of angular
momentum into orbital and spin parts. On the other hand, one could
claim that the $(0j)$ components of spin do not have any physical
meaning and should not appear at all as nonzero quantities. We could
then postulate them to vanish. In the case of a Dirac field, such a rule
would make the choice of a Lagrangean perfectly unique, leaving us
with $\mathcal{L}_{F0}$ (\ref{LF0}). Although the rule may seem to be rather
artificial, it is an example of how one can try to lower the degree of
arbitrariness incorporated in $EC$ theory. It could be interesting
to test it for other types of matter fields. 

Another possible restriction of the freedom of choice of Lagrangean
density could be to require the spin tensor to have as few independent
components as possible. In general, $S^{\mu\nu\rho}=-S^{\nu\mu\rho}$
represents $4\times 6=24$ independent components. If (\ref{LF0}) is
chosen as a Lagrangean density, the spin tensor appears to be totally
antisymmetric, thus having only $4$ independent components. Hence, this
criterion would again distinguish $\mathcal{L}_{F0}$ as an appropriate
Lagrangean density, in a unique manner.

\section{Einstein--Cartan gravity}\label{ECgravity}

\subsection{Field equations}
The Lagrangean four--form of the theory is $\mathfrak{L}=\mathfrak{L}_G+\mathfrak{L}_m$ , where 
$\mathfrak{L}_G=-\frac{1}{4k}\epsilon_{abcd}e^a\wedge e^b\wedge
\Omega^{cd}$ represents gravitational part and $\mathfrak{L}_m$ the matter
part. Here $k=8\pi G$, where $G$ is gravitational constant, $e^a=e^a_{\mu}\d x^{\mu}$ is an
orthonormal cotetrad, ${\omega^a}_b={\Gamma^a}_{bc}e^c$ are connection one--forms
(spin connection) obaying the antisymmetry condition
$\omega_{ab}=-\omega_{ba}$ and
${\Omega^a}_b:=\d{\omega^a}_b+{\omega^a}_c\wedge{\omega^c}_b=\frac{1}{2}{R^a}_{bcd}e^c\wedge
e^d$ the
curvature two--forms. The connection coefficients ${\Gamma^a}_{bc}$
are related to the metric connection $\nabla$ on RC manifold by
$\nabla_{\tilde{e}_c}\tilde{e}_b={\Gamma^a}_{bc}\,\tilde{e}_a$, where
$\tilde{e}_a=e^{\mu}_a\partial_{\mu}$ is an orthonormal tetrad
(a basis of vector fields which is dual to one--form field basis
$e^a$). Variation is given by
\[
\delta{\mathfrak{L}}=\delta e^a\wedge
\l{\frac{\delta\mathfrak{L}_G}{\delta e^a}+\frac{\delta\mathfrak{L}_m}{\delta e^a}}\r+
\delta\omega^{ab}\wedge
\l{\frac{\delta\mathfrak{L}_G}{\delta\omega^{ab}}+\frac{\delta\mathfrak{L}_m}{\delta\omega^{ab}}}\r+
\delta\phi^A\wedge\frac{\delta\mathfrak{L}_m}{\delta\phi^A} \ ,
\]
$\phi^A$ representing matter fields (we used the independence of
$\mathfrak{L}_G$ on $\phi^A$). Explicitly,
\[
\frac{\delta\mathfrak{L}_G}{\delta e^a}=-\frac{1}{2k}\epsilon_{abcd}
e^b\wedge\Omega^{cd} \ , \quad 
\frac{\delta\mathfrak{L}_G}{\delta\omega_{ab}}=-\frac{1}{2k}{\epsilon_{cd}}^{ab}Q^c\wedge
  e^d \ ,
\]
where $Q^a:=De^a=\frac{1}{2}{T^a}_{bc}e^b\wedge e^c$ is a torsion
two--form, whose components in a tetrad basis we have denoted by
${T^a}_{bc}$. The field equations are
\be\label{FEq}
\left. 
\begin{array}{ccc}
\dfrac{\delta\mathfrak{L}_G}{\delta e^a}+\dfrac{\delta\mathfrak{L}_m}{\delta e^a}=0 \hfill & \quad \Leftrightarrow \quad &
{G^a}_b:={R^a}_b-\frac{1}{2}R\delta^a_b=k\,{{\tilde t}\,_b}^a \hfill \\
  &   &   \\
\dfrac{\delta\mathfrak{L}_G}{\delta\omega^{ab}}+\dfrac{\delta\mathfrak{L}_m}{\delta\omega^{ab}}=0 \hfill & \quad \Leftrightarrow \quad &
T^{cab}-T^a\eta^{bc}+T^b\eta^{ac}=k\tilde{S}^{abc}  \hfill \\
\end{array}
\right.
\ee
where ${R^a}_b:=\eta^{ac}{R^d}_{cdb}$, $R:={R^a}_a$, $T^a:={T^{ba}}_
b$ and the dynamical definitions of energy--momentum and spin density tensors on
Riemann--Cartan space (for calculational convenience each of them given below in
two equivalent forms) are
\be\label{ts}
{\tilde t}_{ab}e^b:=-\star\dfrac{\delta\mathfrak{L}_m}{\delta e^a} \ \ \equiv \ \
{{\tilde t}\,_a}^b\epsilon:=\dfrac{\delta\mathfrak{L}_m}{\delta e^a}\wedge e^b \ , \qquad
{\tilde S}^{abc}e_c:=2\star\dfrac{\delta\mathfrak{L}_m}{\delta\omega_{ab}} \ \ \equiv \ \ 
\frac{1}{2}{\tilde S}^{abc}\star e_c:=\dfrac{\delta\mathfrak{L}_m}{\delta\omega_{ab}} \ .
\ee

\subsection{Compatibility between Noether and dynamical definitions of spin density and energy--momentum tensors}

It seems also natural to promote to energy--momentum and spin density
tensors the expressions obtained from Noether currents
(\ref{t}), (\ref{S}) via MCP. We will
denote the resulting tensors by ${t_a}^b$ and $S^{abc}$ (the
corresponding objects in
(\ref{ts}) were denoted with $\tilde{}$ to distinguish from what we
consider now). There is no reason in general for 
${t_a}^b$ and ${S}^{abc}$ to be equal to $\tilde{t_a}^b$ and
$\tilde{S}^{abc}$. Perhaps one could demand such equality to hold and
thus restrict the freedom of choice of the flat space Lagrangean and
the resulting EC theory? 

When the transformation (\ref{Lch}) is applied to the flat space
Lagrangean, the Lagrangean four--form on RC manifold obtained by
standard MCP transforms as
\[
\mathfrak{L}'=\mathfrak{L}-\star e_a\wedge \d
V^a-\star e_a\wedge {\omega^a}_b V^b \ .
\] 
This induces transformation rules for energy--momentum and spin
density tensor components
\be\label{tSgrch}
\ba
&{{\tilde t}\,'_a}^b={{\tilde t}\,_a}^b+\nabla_a V^b-\delta^b_a\,\nabla_c V^c \ , \\
&{{\tilde S}\,'\,}^{abc}={\tilde S}^{abc}+\eta^{ac}V^b-\eta^{bc}V^a \ .
\ea
\ee
Comparison with (\ref{tSch}) allows to conclude that
$
{t_a}^b={{\tilde t}\,_a}^b \ \Leftrightarrow \ {t\,'_a}^b={{\tilde t}\,'_a}^b  .
$
Hence, the two definitions of energy--momentum tensor give the
same result either for all Lagrangeans related by the equivalence relation (\ref{Lch}) or for none
of them. In the case of a Dirac field, the equivalence class
defined by (\ref{LF0}) appears to work well. In particular, if
gravity is minimally coupled to (\ref{LF0}) itself, the resulting
Lagrangean four--form is
\be\label{grLF0}
\left. 
\begin{array}{c}
\hfill \tilde{\mathfrak{L}}_{F0}=
-\dfrac{i}{2}\star e_a\wedge \l{ \ov{\psi}\gamma^a
  D\psi-\ov{D\psi}\gamma^a\psi}\r-m\ov{\psi}\psi\,\epsilon \ , \hfill \\
 \\
\hfill D\psi:=\d\psi-\dfrac{1}{2}\omega_{ab}\Sigma^{ab}\psi \ , \quad 
\ov{D\psi}=\l{D\psi}\r^{\dagger}\gamma^0  \hfill \\
\end{array}
\right.
\ee
and the energy--momentum tensor is
\[
{t_a}^b={{\tilde t}\,_a}^b=\frac{i}{2}\left({\ov{\psi}\gamma^b\nabla_a\psi-\ov{\nabla_a\psi}\,\gamma^b\psi}\right)-
\delta^b_a 
\left[{\frac{i}{2}\left({\ov{\psi}\gamma^c\nabla_c\psi-\ov{\nabla_c\psi}\,\gamma^c\psi}\right)-m\ov{\psi}\psi}\right]
\]
(here $\nabla_a\psi$ are components of a one--form $D\psi$ in the
cotetrad basis: $D\psi=(\nabla_a\psi)\,e^a$ ). Note that the
cannonical energy--momentum tensor is the one which appears in the
`Einstein equation' (\ref{FEq}), not the symmetric one obtained by
Belinfante--Rosenfeld method.

In the case of spin, we will confine our considerations to the Dirac field
and to the vector field of the form (\ref{VB}). For (\ref{LF0}), we
find by straightforward calculations based on (\ref{grLF0}), (\ref{ts}) and (\ref{S}) that
\[
S^{abc}=\tilde{S}^{abc}=\frac{1}{4}\ov{\psi}\,\left\{{\Sigma^{ab},\gamma^c}\right\}\,\psi \ .
\]
Then, for the Lagrangean density $\mathcal{L}_{F0}+\partial_{\mu}V^{\mu}$, we
get from (\ref{Sch}) and (\ref{tSgrch})
\[
\tilde{S}'\,^{abc}-S'\,^{abc}=\ov{\psi}
\l{\,\frac{i}{2}[\Sigma^{ab},B^c]-\eta^{ac}B^b+\eta^{bc}B^a \,}\r \psi \ ,
\]
which vanishes identically on account of
$[\gamma^a,\Sigma^{bc}]=4i\eta^{a[b}\gamma^{c]}$. Hence, both definitions are perfectly compatible, independently
of the choice of flat Lagrangean from the equivalence class of
(\ref{LF0}) (the equivalence relation being given by (\ref{Lch})). It is worth noting that in the case of gauge fields such
compatibility would not occur, unless we use MCP in a naive
manner, $F_{\mu\nu}=\partial_{\mu}A_{\nu}-\partial_{\nu}A_{\mu}+[A_{\mu},A_{\nu}]\longrightarrow
\nabla_{\mu}A_{\nu}-\nabla_{\nu}A_{\mu}+[A_{\mu},A_{\nu}]$, which breaks gauge symmetry.

\section{Effective action, modified Dirac equation and their physical meaning}\label{effective}
\subsection{Derivation of the effective action}\label{derivation}

Let us define the {\it contorsion one--forms}
\[
{K^a}_b={K^a}_{bc}\,e^c:={\omega^a}_b-\stackrel{\circ}{\omega}{{^a}_b}
\]
(objects with $\circ$ above will always denote torsion--free objects,
related to LC connection). The curvature two--form decomposition
\[
{\Omega^a}_b=\stackrel{\circ}{\Omega}{^a}_b+{\stackrel{\circ}{D}\,}{K^a}_b+{K^a}_c\wedge{K^c}_b
\]
results in
\beq
\ba
&\mathfrak{L}_G:=-\frac{1}{4k}\,\epsilon_{abcd}\,e^a\wedge e^b\wedge{\Omega^{cd}}
={\stackrel{\circ}{\mathfrak{L}}_G}-\frac{1}{4k}\epsilon_{abcd}\,e^a\wedge e^b\wedge{K^c}_e\wedge{K^{ed}}
-\frac{1}{4k}{\,\stackrel{\circ}{D}}\l{\,\epsilon_{abcd}\,e^a\wedge e^b\wedge{K^{cd}}\,}\r \ ,
\ea
\eeq
where ${\stackrel{\circ}{D}}\epsilon_{abcd}=0$ was used.
Here $k=8 \pi G$, where
$G$ is a gravitational constant. Since all Lorentz indexes in the
last term are contracted, ${\stackrel{\circ}{D}}$ acts like a usual differential. 
Using the relation between components of contorsion and torsion
tensors
\beq
K_{abc}=\frac{1}{2}(T_{cab}+T_{bac}-T_{abc})
\eeq
and decomposing torsion into its irreducible parts
\beq
\ba
T_{abc}=\frac{1}{3}(\eta_{ac}T_b-\eta_{ab}T_c)+\frac{1}{6}\epsilon_{abcd}S^d+q_{abc}
\ , \qquad
T_a:={T^b}_{ab} \ , \quad S_a:=\epsilon_{abcd}T^{bcd} \ ,
\ea
\eeq
we can finally obtain 
\beq
\mathfrak{L}_G={\stackrel{\circ}{\mathfrak{L}}_G}+\frac{1}{2k}
\l{\frac{2}{3}T_aT^a-\frac{1}{24}S_aS^a- 
\frac{1}{2}q_{abc}q^{abc}}\r \, 
\epsilon + \d (\dots) 
\eeq
(the last term is a differential whose particular form will not be
needed). Similarly, the Dirac Lagrangean (\ref{grLF0}) decomposes as
\beq
\tilde{\mathfrak{L}}_{F0}= 
\stackrel{\circ}{\tilde{\mathfrak{L}}}_{F0}
-\frac{1}{8}S_aJ^a_{(A)} \, 
\epsilon \ .
\eeq
The addition of a divergence of a vector field $V$ to the flat space
Lagrangean results in one more term (\ref{TV}). Ultimately, we have the following four--form on
RC space representing EC gravity with fermions
\be\label{Ltot}
\ba
\mathfrak{L}={\stackrel{\circ}{\mathfrak{L}}_G}+
\stackrel{\circ}{\tilde{\mathfrak{L}}}_{F0}+ \d (\dots)+
\frac{1}{2k}\l{\frac{2}{3}T_aT^a-\frac{1}{24}S_aS^a-\frac{1}{2}q_{abc}q^{abc}-
\frac{k}{4}S_aJ^a_{(A)}-2kT_aV^a}\r\,\epsilon \ .
\ea
\ee
Variation of the resulting action with respect to $T_a$, $S_a$ and $q_{abc}$
yields the equations
\be\label{TSV}
\ba
T^a=\frac{3k}{2} V^a \ , \qquad S^a=-3kJ^a_{(A)} \ , \qquad q_{abc}=0 \ .
\ea
\ee
Inserting these results into (\ref{Ltot}), we finally get the effective
Lagrangean four--form
\be\label{ECeff}
\mathfrak{L}_{eff}={\stackrel{\circ}{\mathfrak{L}}_G}+\stackrel{\circ}{\tilde{\mathfrak{L}}}_{F0}+
\frac{3k}{16}\l{J^{(A)}_aJ_{(A)}^a-4V_aV^a}\r\,\epsilon \ .
\ee
The total differential has been omitted in the final formula. 
Note that beyond the well--known axial--axial interaction term
\cite{HH} we have
an additional one due to the ambiguity (\ref{Lch}). In the case of $V$ being a linear combination of axial and vector
currents (\ref{Vab}) we have 
\be\label{ECeffab}
\mathfrak{L}_{eff}={\stackrel{\circ}{\mathfrak{L}}_G}+ 
\stackrel{\circ}{\tilde{\mathfrak{L}}}_{F0}+
\l{C_{AA} \ J^{(A)}_aJ_{(A)}^a+C_{AV} \ J^{(A)}_aJ_{(V)}^a+C_{VV} \ J^{(V)}_aJ_{(V)}^a}\r\,\epsilon \ ,
\ee
\be\label{ECc}
C_{AA}=\frac{3k}{16}(1-4b^2) \ , \qquad
C_{AV}=-\frac{3k}{2}ab \ , \qquad
C_{VV}=-\frac{3k}{4}a^2 \ , 
\ee
where $a,b$ are completely arbitrary real numbers! Hence, in the most
generic case we
have three types of possible contact interactions and we cannot claim
that none of them is small. As far as we cannot eliminate the
ambiguity (\ref{Lch}), there is nothing on the basis of which
their smallness could be conjectured. We can only establish
experimentally some bounds on the values of $a$ and
$b$. To see how this could possibly be done, observe that the
nonlinear equation for $\psi$
\be\label{dir}
\l{i\,\gamma^a{\stackrel{\circ}{\nabla}}_a-m}\r\psi+
\left[{-2C_{AA} J^{(A)}_a\gamma^5+C_{AV}\l{J^{(A)}_a-J^{(V)}_a\gamma^5}\r+2C_{VV}J^{(V)}_a}\right] 
\gamma^a\psi=0 
\ee
obtained from (\ref{ECeffab}) via variational procedure does not reduce to the usual Dirac equation in the limit of vanishing
Riemannian curvature. For the space--time metric being flat, the
first term of (\ref{dir}) reduces to the usual Dirac one, but the
remaining two preserve their forms. One could try to interpret
physically the resulting equation
\cite{Ker}\cite{HD}\cite{Rumpf}\cite{CI}, aiming ultimately at
measuring physical effects produced by $a$ and $b$ by flat space experiments.

It is worthy to make two further observations concerning EC theory
with fermions. Firstly, a parity violating term in (\ref{ECeffab})
appeared. Secondly, for $V=\frac{1}{2}J_{(A)}$ ($a=0,b=\frac{1}{2}$) interaction terms in
(\ref{ECeff}) cancel out and the effective action
of EC theory appears to be the same as the usual GR one. However, we
should not think prematurely of these two theories as being
indistinguishable, since  the EC one introduces nonvanishing torsion on
space--time, $T^a=\frac{3k}{4}J_{(A)}^a$, $S^a=-3kJ_{(A)}^a$, as
follows from (\ref{TSV}) for $V=\frac{1}{2}J_{(A)}$. As
for parity violation, one would avoid it by demanding
the flat space Dirac Lagrangean density taken as starting point to be parity
invariant. This would result in $V\sim J_{(V)}$ ($b=0$). In this case,
(\ref{ECeff}) is parity invariant and contains two interaction terms
with axial--axial interaction having the fixed, well--known, small coupling constant
and vector--vector one having unknown coupling constant (possibly high
enough to be
measurable in the near future). One could argue that the case $V\sim
J_{(V)}$ is what we should expect, since
introduction of $V$ having axial or mixed axial--vector transformation
properties seems rather unnatural. Although the effective action
is parity invariant both in the case of $V$ being vector, as well as an
axial vector (but not their combination), the first equation of
(\ref{TSV}) will not violate parity if and only if $V$ is a vector
(since $T^a$ is).

\subsection{Is it possible to distinguish between GR and EC by
  measuring the strength of fermion interactions?}

In the previous section we have pointed out that we
could choose between EC theory and GR by torsion measurements. However, it seems very difficult to measure torsion
directly \cite{Sh}. There is another, much more promising possibility. First of
all, note that the Lagrangean four--form (\ref{ECeffab}) is relevant for
both GR and EC theory, but for GR all coupling constants $C_{AA},C_{AV},C_{VV}$ vanish, whereas
for EC theory they are given by (\ref{ECc}).
These constants are much more likely to be measurable in practice than
torsion, as they are responsible for the strength of corresponding
point interactions between fermions. Let us imagine that we can separate the
contributions coming from different types of these interactions in experiments
that we perform. As long as we get values of all constants indistinguishable
from zero, we are not able to say which of the two theories of
gravitation is correct. Measuring a non--zero value of at least one of
them would provide an argument against standard GR. Of course, for any
values of $C$'s, one could produce the effective Lagrangean
(\ref{ECeffab}) on the base of the torsionless approach of standard GR
by simply adopting from the beginning 
\be\label{flateff}
\mathcal{L}_{Fab}={\mathcal{L}_{F0}}+
C_{AA} \ J^{(A)}_aJ_{(A)}^a+C_{AV} \ J^{(A)}_aJ_{(V)}^a+C_{VV} \ J^{(V)}_aJ_{(V)}^a
\ee
as flat space Lagrangean density for fermions. However, on the ground
of EC theory the interaction terms arise naturally 
as a necessary consequence of the relation between torsion and matter, as explained
in Subsection  \ref{derivation}. In this paper, we aim to treat both
the theories in the most natural manner, adopting the simplest Dirac
theory of fermions in flat space as a starting point in each case. Then
the most traditional method of minimal coupling is used to incorporate
gravity and the results are compared. According to this standpoint, a non--zero value of a coupling
constant would discredit GR. This would not necessarily mean
the confirmation of EC theory. Even in the case of generic values of
$a$ and $b$, such result could
either agree with EC or contradict it. The first possibility realises if
the set of equations (\ref{ECc}) have a solution with respect to $a$
and $b$ for measured values of $C's$, whereas the second one
corresponds to the alternative case. Hence, the
number of free parameters appearing in EC theory with fermions is
smaller than the number of independent parameters whose values can be
experimentally established. The measurements of interactions between
fermions can then suffice either to confirm or to contradict the theory.

\subsection{The background field approximation}

It seems extremely difficult to extract any information about the
physical nature of a gravity--induced fermion interaction without some
simplifying assumptions. The background field approach suggests
\cite{Rumpf}\cite{Ker} that
in the case of (\ref{LF0}) being the flat space Lagrangean ($a=b=0$), the
interaction is repulsive for particles with aligned spins, attractive
for antiparallel spins and universal (independent on whether particles
or antiparticles are considered). Let us investigate what will change
if possibility of divergence addition to the flat Lagrangean is taken
into account. We shall assume that the test Dirac particle $\psi$ has
a negligible influence on space--time torsion, while compared to that
of a background Dirac field $\psi_{bg}$. Hence, (\ref{Ltot}) is still
valid, but whereas $\psi$ appearing in $\stackrel{\circ}{\tilde{\mathfrak{L}}}_{F0}$, $J_{(A)}$
and $V$ is the test field whose dynamics we
wish to describe, the torsion components are totally determined by
$\psi_{bg}$ via the formulas (\ref{TSV}). For both the test and
background field, the relation $V=aJ_{(V)}+bJ_{(A)}$ holds for some
real numbers $a$ and $b$ (necessarily the same in both cases, in order
for our approach to be logically consistent -- according to our
viewpoint, the values of these constants are determined by the most
reasonable choice of flat space Lagrangean for fermions). Explicitly,
the Lagrangean four--form is now
\be\label{Lbg}
\ba
\mathfrak{L}_{bgApp}=\stackrel{\circ}{\mathfrak{L}}_G
-\frac{3k}{16}\l{J^{(A)bg}_aJ_{(A)bg}^a-4V^{bg}_aV_{bg}^a}\r\,\epsilon
+\stackrel{\circ}{\tilde{\mathfrak{L}}}_{F0} 
+\frac{3k}{8}\l{J^{(A)bg}_aJ_{(A)}^a-4V^{bg}_aV^a}\r\,\epsilon
\ea
\ee
(for total differential discarded). Only the two last terms depend on
the test field and variation with respect to it yields a {\bf linear}
equation for $\psi$. It looks the same as (\ref{dir}) with
$J^{(A)}$, $J^{(V)}$ replaced by $J^{(A)bg}$, $J^{(V)bg}$. In the
limit of the space--time metric being flat, it
can be rewritten in the form
\be\label{Hdir}
i\partial_t\psi=H\psi
\ee
for the Hamilton operator
\beq
\ba
&H=-i\gamma^0\gamma^j\partial_j+\tilde{H} \ , \quad 
\tilde{H}=m\gamma^0-
\left[{2C_{AA}J^{(A)bg}_a\gamma^5+C_{AV}\l{J^{(A)bg}_a+J^{(V)bg}_a\gamma^5}\r+2C_{VV}J^{(V)bg}_a}\right]\gamma^0\gamma^a \ ,
 \quad j=1,2,3 \ .
\ea
\eeq

In the following we will refer to particles and antiparticles as
electrons and positrons. From now on we
shall adopt the Dirac representation for $\gamma$'s 
\beq
\gamma^0 =
\l \begin{array}{ccc}
{\bf 1} & 0 \\
0 & -{\bf 1} \\
\end{array} \r \ , \quad
\gamma^j =
\l \begin{array}{ccc}
0 & \sigma^j \\
-\sigma^j & 0 \\
\end{array} \r \ , \quad
\gamma^5 =-i\gamma^0\gamma^1\gamma^2\gamma^3=-
\l \begin{array}{ccc}
0 & {\bf 1} \\
{\bf 1} & 0 \\
\end{array} \r \ ,
\eeq
where $\sigma^j$ are Pauli matrixes. Following \cite{Rumpf} and
\cite{Ker}, we shall consider the background Dirac field of the form
\beq
\psi_{bg\uparrow}=\sqrt{n} \ e^{-i\,m\,t}
\l \begin{array}{c}
1 \\
0 \\
0 \\
0 \\
\end{array} \r \ , \quad n\in\mathbb{R}_{+} \ ,
\eeq
which simulates an `electron distribution of number density
$n$' \cite{Ker} with their spins directed upwards along a fixed axes
of quantisation. We will consider test particles at rest,
$-i\partial_j\psi=0$, for which
only $\tilde{H}$ part of $H$ is important. For the postulated
background it equals
\beq
\ba
\tilde{H}=
\l\begin{array}{cccc}
m+2n(C_{AA}-C_{VV})&0&0&0\\
0&m-2n(C_{AA}+C_{VV})&0&2n C_{AV}\\
0&0&-m+2n(C_{AA}-C_{VV})&0\\
0&2nC_{AV}&0&-m-2n(C_{AA}+C_{VV})\\
\end{array}\r \ .
\ea
\eeq
Let us recall that the theory does not break parity if and only if $b=0$
(see (\ref{TSV})). In this case
$C_{AA}=\frac{3k}{16}$ and $C_{VV}=-\frac{3k}{4}a^2$ and the vectors
\beq
\psi_{\uparrow}=e^{-iE_{\uparrow}t}
\l \begin{array}{c}
1 \\
0 \\
0 \\
0 \\
\end{array} \r \ , \quad
\psi_{\downarrow}=e^{-iE_{\downarrow}t}
\l \begin{array}{c}
0 \\
1 \\
0 \\
0 \\
\end{array} \r \ , \quad
\tilde{\psi}_{\uparrow}=e^{-i\,\tilde{E}_{\uparrow}t}
\l \begin{array}{c}
0 \\
0 \\
1 \\
0 \\
\end{array} \r \ , \quad
\tilde{\psi}_{\downarrow}=e^{-i\,\tilde{E}_{\downarrow}t}
\l \begin{array}{c}
0 \\
0 \\
0 \\
1 \\
\end{array} \r \ , \quad
\eeq
provide the set of eigenvectors of $\tilde{H}$. They solve equation
(\ref{Hdir}) for $E_{\uparrow}=m+\frac{3}{8}kn(1+4a^2) \ $, $ \
E_{\downarrow}=m-\frac{3}{8}kn(1-4a^2) \ $,
$ \ \tilde{E}_{\uparrow}=-m+\frac{3}{8}kn(1+4a^2) \ $, $ \
\tilde{E}_{\downarrow}=-m-\frac{3}{8}kn(1-4a^2) \ $ being the 
corresponding eigenvalues of $\tilde{H}$. Hence, in the case of $|a|<\frac{1}{2}$, the
energies are shifted upwards for aligned spins and downwards for
antiparallel spins, which agrees with the conclusions of
\cite{Ker}\cite{Rumpf} concerning attractivity and repulsivity, as
well as universality of gravity--induced fermion interactions. Note, however, that the magnitude of interaction is
different for aligned and antiparallel spins. For $|a|=\frac{1}{2}$
the antiparallel spins seem not to interact at all, whereas for
$|a|>\frac{1}{2}$ all interactions are repulsive. It is important to
note that the independence of results from whether particles or
antiparticles are considered remains preserved in all cases, which
allows us to believe in a genuine gravitational nature of the
investigated phenomenon. One can easily find out that our conclusions
do not change under the replacement of $\psi_{bg\uparrow}$ by
$\psi_{bg\downarrow}$, $\tilde{\psi}_{bg\uparrow}$ or
$\tilde{\psi}_{bg\downarrow}$, which is an indispensable consistency
test of our reasoning.

If $b\not=0$, we have necessarily $C_{AV}\not=0$ and parity
symmetry is broken. Although the third spatial component of spin
operator $S^3=\frac{1}{2}\Sigma^{12}$ still commutes with $\tilde{H}$
and one can construct the basis of $\mathbb{C}^4$ from common
eigenvectors of $\tilde{H}$ and $S^3$, it is difficult to say which of
them describe electrons and which correspond to positrons, as they are
not in general of the form 
$
\l \begin{array}{c}
\kappa \\
0 \\
\end{array} \r
$
or
$
\l \begin{array}{c}
0 \\
\chi \\
\end{array} \r
$
in Dirac representation, nor are their energies of the form 
$m+\delta E$ or $m-\delta E$, for $\delta E$ independent from $m$.

It should be stressed that the background field approach is an
approximate technique, not only because of neglecting the influence of
test field on space--time geometry and all matter fields on the
space--time metric, but also because the Dirac field was not second
quantised, which could signifficantly change our conclusions as it
does in the standard $a=b=0$ case \cite{CI}.

\section{Conclusions}\label{section4}
Minkowski space Lagrangean densities differing by divergence give rise
to generically nonequivalent theories, when Einstein--Cartan gravity is
minimally included. Hence, it is important to choose carefully among
equivalent flat space Lagrangeans. In the case of the Dirac field, a very
natural requirement for Lagrangean density to be real, invariant under
proper Lorentz transformations and not dependent on higher
derivatives or higher powers of fields, leads to the
two--real--parameter family (\ref{ECeffab}) of EC theories for gravity with fermions. There are
some plausible arguments in favour of choosing particular values of the
parameters (namely $a=b=0$), motivated by the resulting form of the spin
density tensor. This
choice leads to the theory discussed in the earlier papers. However,
these arguments are not completely convincing. Rejecting them, we
are obliged to take the entire two--parameter family seriously. Then,
the gravity--induced fermion interaction acquires a three--fold structure described
by the axial--axial, vector--vector and axial--vector term appearing in
the effective action (\ref{ECeffab}). The strength of these constituent parts of the
total interaction is governed
by the coupling constants, which depend on the parameters of the
theory (\ref{ECc}). Such values of the parameters can be chosen that the interaction do not occur
at all and thus EC theory is indistinguishable from the standard GR on the
level of effective action. The theories still differ by the presence
of nonzero torsion in EC theory. A reasonable requirement for the flat
fermionic Lagrangean density to be parity invariant reduces the number
of parameters to one. The resulting EC theory is then parity invariant and
differs from GR on the effective level. Although the axial--axial
constituent of the fermion interaction has now a well--known, small coupling constant, the
strength of the vector--vector part of the interaction is not restricted and may
achieve significant values, possibly leading to the effects observable in
practice. The background field approximation analysis suggests that
the vector--vector constituent of the interaction is repulsive, for both
aligned and antiparallel spins of particles. If significantly
strong, it will dominate over the axial--axial part, making all
fermions to repulse each other. Hence, the possibility of modifying EC
theory with fermions by the freedom of adding a divergence to the flat fermionic
Lagrangean can lead to the meaningful, interesting and new physical
effects that should not be disregarded until some reasonable procedure
for restricting this freedom is proposed and commonly accepted.

\section*{Acknowledgements}
I would like to thank Wojciech Kaminski, Ryszard Kostecki and
Urszula Pawlik for stimulating discussions, as well as the help with
{\LaTeX}  and linguistic corrections.

\section{Appendix: Notation and conventions}\label{A1}

Throughout the paper $a,b,\dots$ are orthonormal tetrad indexes and $\mu,\nu,\dots$ correspond to
a holonomic frame. For inertial frame of flat Minkowski space,
which is both holonomic and orthonormal, we use $\mu,\nu,\dots$.
The metric components in an orthonormal tetrad basis $\tilde{e}_a$ are
 $g\l{\tilde{e}_a,\tilde{e}_b}\r=(\eta_{ab})=diag(1,-1,-1,-1)$. Lorentz
 indeses are shifted by $\eta_{ab}$. $\epsilon=e^0\wedge e^1\wedge
e^2\wedge e^3$ denotes the cannonical
volume four--form  whose components in orthonormal tetrad basis obey $\epsilon_{0123}=-\epsilon^{0123}=1$.
The action of a covariant exterior differential $D$ on any $(r,s)$-tensorial
type differential $m$--form 
\beq
{T^{a_1\dots a_r}}_{b_1\dots b_s}=
\frac{1}{m!}{T^{a_1\dots a_r}}_{b_1\dots b_s\mu_1\dots \mu_m}\d x^{\mu_1}\w\dots\w\d x^{\mu_m}
\eeq
is given by 
\beq
D{T^{a_1\dots a_r}}_{b_1\dots b_s}:=\d {T^{a_1\dots a_r}}_{b_1\dots b_s}+
\sum_{i=1}^r{\omega^{a_i}}_c\w {T^{a_1\dots c\dots a_r}}_{b_1\dots b_s}-
\sum_{i=1}^s{\omega^c}_{b_i}\w {T^{a_1\dots a_r}}_{b_1\dots c\dots b_s} \ .
\eeq
The hodge star action on external products of orthonormal cotetrad
one--forms is given by
\beq
\star e_a=\frac{1}{3!}\epsilon_{abcd}e^b\w e^c\w e^d \ , \quad 
\star \l{e_a\w e_b}\r=\frac{1}{2!}\epsilon_{abcd}e^c\w e^d \ , \quad 
\star \l{e_a\w e_b\w e_c}\r=\epsilon_{abcd}e^d \ ,
\eeq
which by linearity determines the action of $\star$ on any differential
form.

\section{Appendix: Noether theorem}

Let $S[\phi^A]=\int\mathcal{L}\l{\phi^A,\partial_{\mu}\phi^A}\r\d^4 x \ $
represent  the action of a field theory in Minkowski space $\mathbb{M}$. Consider a
Lie group $\mathcal{G}$ which acts on space--time, as well as a target
space $\mathbb{T}$ in which the fields $\phi^A$ take their values, as a group of transformations. Let 
\be\label{symtr}
\ba
x^{\mu}\longrightarrow x'^{\mu}=x^{\mu}+\delta x^{\mu} \ , \\
\phi^A\longrightarrow \phi'^A=\phi^{A}+\delta \phi^{A} 
\ea
\ee
represent the infinitesimal form of the action of $\mathcal{G}$ on
$\mathbb{M}$ and $\mathbb{T}$ respectively. The transformations are
{\it symmetry transformations} of the theory if they do not change the
action, up to possibly surface terms (and thus leave the form of field
equations invariant). This is equivalent to the condition
\be\label{symm}
\mathcal{L}\l{\phi'^A,\partial'_{\mu}\phi'^A}\r\d^4x' = 
\mathcal{L}\l{\phi^A,\partial_{\mu}\phi^A}\r\d^4x+\partial_{\mu}W^{\mu}\d^4x
\ee
for some vector field $W$. Note that in our approach, $\phi^A$
transform only under the action of $\mathcal{G}$ in target space --
the coordinates $x^{\mu}$ `hidden' in $\phi^A$ do not undergo any
transformation. For infinitesimal transformations we have
\beq
\ba
&\mathcal{L}\l{\phi'^A,\partial'_{\mu}\phi'^A}\r\approx\mathcal{L}\l{\phi^A,\partial_{\mu}\phi^A}\r
+\frac{\partial\mathcal{L}}{\partial\phi^A}\delta\phi^A+ 
\frac{\partial \mathcal{L}}{\partial(\partial_{\mu}\phi^A)}
\l{\partial_{\mu}\delta\phi^A-\partial_{\nu}\phi^A\partial_{\mu}\delta  x^{\nu}}\r \ , \\
&\d^4x'\approx\d^4x+\partial_{\mu}\delta x^{\mu}\d^4x 
\ea
\eeq
and (\ref{symm}) appears to be equivalent to 
\be\label{symcond}
\frac{\partial\mathcal{L}}{\partial\phi^A}\delta\phi^A+ 
\frac{\partial \mathcal{L}}{\partial(\partial_{\mu}\phi^A)}
\l{\partial_{\mu}\delta\phi^A-\partial_{\nu}\phi^A\partial_{\mu}\delta x^{\nu}}\r+\mathcal{L}\partial_{\mu}\delta x^{\mu}
=\partial_{\mu}W^{\mu}
\ee
which can be finally expressed in the form
\beq
\partial_{\mu}j^{\mu}=
\l{\frac{\partial\mathcal{L}}{\partial\phi^A}-\partial_{\mu} 
\frac{\partial \mathcal{L}}{\partial(\partial_{\mu}\phi^A)}}\r
\l{\partial_{\nu}\phi^A\delta x^{\nu}-\delta\phi^A}\r \ ,
\eeq
where 
\beq
\ba
&j^{\mu}=
-{t_{\nu}}^{\mu}\delta x^{\nu}+\frac{\partial\mathcal{L}} 
{\partial(\partial_{\mu}\phi^A)} 
\delta\phi^A-W^{\mu} \ , \\
&{t_{\nu}}^{\mu}= 
\frac{\partial
  \mathcal{L}}{\partial(\partial_{\mu}\phi^A)} 
\partial_{\nu}\phi^A 
-\delta^{\mu}_{\nu}\mathcal{L}
\ea
\eeq
is a Noether current associated to the symmetry transformation (\ref{symtr}), which
is clearly conserved, i.e. $\partial_{\mu}j^{\mu}=0$, if the
Euler--Lagrange equations for fields are satisfied. Note that in the case
of transformations which do not change the volume--form $\d^4x$,
such as space--time translations and Lorentz transformations, the
condition for them to be symmetries (\ref{symm}) is fulfilled if and
only if the Lagrangean density changes by a divergence only. For simplicity,
in Section \ref{EMS} of this paper we consider only the situation when
$\mathcal{L}$ is left just invariant by Lorentz transformations
(invariance under space--time translations is obvious, since
Lagrangean densities under consideration do not depend explicitly on
$x$). In particular, this is the case for (\ref{LF0}) and all
Lagrangean densities related to (\ref{LF0}) via (\ref{Lch}). The
possibility of construction of Lagrangean densities violating such
invariance by the addition of a divergence of a vector field having weird
transformation properties is not interesting in the context of this
paper. What we wish to do is to limit the multiplicity of
equivalent flat space Lagrangean densities. The requirement of Lorentz
invariance should be understood as the first restriction that we
impose on them in order to find the most appropriate one.

Some references (e.g. \cite{Wein}) adopt a different approach, in which the derivatives $\partial_{\mu}$ and measure
$\d^4x$ remain untouched by the transformation. Instead, $x^{\mu}$
that is hidden in the fields $\phi^A$ does transform. Hence,
$\phi^A\l{x^{\mu}}\r$ transforms under the combined action of
$\mathcal{G}$ on both $\mathbb{M}$ and $\mathbb{T}$:
\beq
\ba
&\phi^A\l{x^{\mu}}\r\rightarrow\phi'^A\l{x'^{\mu}}\r\approx\phi^A\l{x^{\mu}-\delta x^{\mu}}\r+\delta\phi^A\l{x^{\mu}-\delta x^{\mu}}\r\approx
\phi^A-\partial_{\mu}\phi^A+\delta\phi^A=\phi^A+\tilde{\delta}\phi^A \ , \\
&\tilde{\delta}\phi^A=\delta\phi^A-\partial_{\mu}\phi^A\delta x^{\mu} \  .
\ea
\eeq
The condition for the transformation to be a symmetry is now
\beq
\frac{\p\LL}{\p\phi^A}\tilde{\delta}\phi^A+\frac{\p\LL}{\p(\p_{\mu}\phi^A)}\,\p_{\mu}\tilde{\delta}\phi^A=
\p_{\mu}\tilde{W}^{\mu} \ .
\eeq
One can easily find this condition to be equivalent to (\ref{symcond})
by putting $\tilde{W}^{\mu}=W^{\mu}-\LL\delta x^{\mu}$. Note however
that Lagrangean densities that remain invariant under Lorentz
transformations according to the first viewpoint acquire an additional
divergence term $\p_{\mu}\l{\LL\delta x^{\mu}}\r$ according to the
second interpretation. This is why some references \cite{Wein} claim
the Poincar\'e transformations to change the simplest Lagrangean
densities of field theory, which is not the case in the approach
adopted here.

\end{document}